# Seasonal variations of chemical species and haze in Titan's upper atmosphere

Siteng Fan[1,2,*], Daniel Zhao[1,3], Cheng Li[4], Donald E. Shemansky[5], Mao-Chang Liang[6], Yuk L. Yung[1,7]

1. Division of Geological and Planetary Sciences, California Institute of Technology, Pasadena, CA 91125, USA
2. LMD/IPSL, Sorbonne Université, PSL Research University, École Normale Supérieure, École Polytechnique, CNRS, Paris 75005, France
3. Department of Mathematics, Harvard University, Cambridge, MA 02138, USA
4. Department of Climate and Space Sciences and Engineering, University of Michigan, Ann Arbor, MI 48109, USA
5. Planetary and Space Science Division, Space Environment Technologies, Pacific Palisades, CA 90272, USA
6. Research Center for Environmental Changes, Academia Sinica, Taipei 115, Taiwan, ROC
7. Jet Propulsion Laboratory, California Institute of Technology, Pasadena, CA 91109, USA

*Corresponding author: stfan@gps.caltech.edu

## ABSTRACT

Seasonal variation is significant in Titan's atmosphere due to the large change of solar insolation resulting from Titan's 26.7° axial tilt relative to the plane of Saturn's orbit. Here we present an investigation of hydrocarbon and nitrile species in Titan's upper atmosphere at 400-1200 km, which includes the mesosphere and the lower thermosphere, over more than one fourth of Titan's year (2006-2014, $L_S$=318°-60°), using eighteen stellar occultation observations obtained by Cassini/UVIS. Vertical profiles of eight chemical species ($CH_4$, $C_2H_2$, $C_2H_4$, $C_2H_6$, $C_4H_2$, $C_6H_6$, HCN, $HC_3N$) and haze particles are retrieved from these observations using an instrument forward model, which considers the technical issue of pointing motion. The Markov-chain Monte Carlo (MCMC) algorithm is used to obtain the posterior probability distributions of parameters in the retrieval, which inherently tests the extent to which species profiles can be constrained. The results show that no change of the species profiles is noticeable before the equinox, while the decrease of atmospheric temperature and significant upwelling in the summer hemisphere are found five terrestrial years afterwards. Altitude of the detached haze layer decreases towards the vernal equinox then it disappears, and no reappearance is identified within the time range of our data, which is consistent with observations from Cassini/ISS. This study provides observational constraints on the seasonal change of Titan's upper atmosphere, and suggests further investigations of the atmospheric chemistry and dynamics therein.

# 1. INTRODUCTION

Titan, the second largest moon in the solar system, has a thick $N_2$-dominant atmosphere with a surface pressure of 1.5 times that on Earth (Lindal et al. 1983). It is tidally locked to Saturn and rotates on an orbit approximately coplanar to Saturn's equator. Due to Saturn's 26.7° obliquity, spatial distribution of solar insolation on Titan has large variations with a 29.5-year orbital period, which leads to strong seasonal variations of Titan's atmosphere. As a result of the long lifetime of the Cassini-Huygens mission (Matson et al. 2002) from 2004 to 2017, investigations of the seasonal variations have become possible. Studies of Titan's lower and middle atmosphere (<500 km), which mostly include the stratosphere, have shown significant changes in atmospheric composition and temperature (e.g., Coustenis et al. 2010, Vinatier et al. 2015, Coustenis et al. 2016, Teanby et al. 2019, Mathé et al. 2020, Vinatier et al. 2020), as well as the altitude of the detached haze layer (West et al. 2011, West et al. 2018, Seignovert et al. 2021). Most of these changes have been well explained by numerical simulations of atmospheric chemistry and dynamics (Yung et al. 1984, Newman et al. 2011, Lebonnois et al. 2012, Lora et al. 2015). In contrast, however, the seasonal change of Titan's upper atmosphere is currently not well understood due to the lack of observational constraints, which is the subject of this study.

The Cassini orbiter and the Huygens lander carry a number of instruments capable of remote sensing and in situ measurements of Titan's atmosphere. Among them, the Ultraviolet Imaging Spectrograph (UVIS, Esposito et al. 2004) mainly covers the upper atmosphere, where the atmospheric organic compositions, hydrocarbons and nitriles, show distinguishable spectral features in ultraviolet (Figure 1, Hörst 2017). Analyses using Cassini/UVIS observations from selected Titan flybys have significantly extended our knowledge about Titan's upper atmosphere during the last decade (Shemansky et al. 2005, Liang et al. 2007, Koskinen et al. 2011, Kammer et al., 2013, Stevens et al. 2015, Capalbo et al. 2016, Fan et al. 2019, Yelle et al. 2021). Vertical profiles of the molecules more complex than methane are constrained by stellar occultation observations in far-ultraviolet (FUV) obtained during six flybys (TB, T21, T41i, T41e, T52, and T53) before Titan's spring equinox in 2009 (Shemansky et al. 2005, Koskinen et al. 2011, Capalbo et al. 2016, Fan et al. 2019), as well as a recent occultation in 2016 of the three stars in Orion's belt (T116, Yelle et al. 2021). However, due to a technical issue, the pointing motion of the spacecraft during occultation measurements (Chiang et al. 1993), most of the observations were not reduced until recently. To address this issue, Fan et al. (2019) proposed a new method that combines a Cassini/UVIS instrument simulator, which considers the spacecraft pointing motion, and the Markov-chain Monte Carlo (MCMC, Mackay 2003) method to constrain species abundances. Vertical profiles of the chemical species were successfully derived during a flyby (T52) with poor spacecraft pointing stability. In this work, we applied the new technique to eighteen stellar occultation observations with decent data quality from 2006 ($L_S$=318°) to 2014 ($L_S$=60°). These observations have a large coverage of season (Figure 2 and Table 1), and therefore enable the investigation of seasonal variations of chemical species in Titan's upper atmosphere.

The remainder of the paper is structured as follows: the methodology of retrieving chemical species abundances is described in Section 2, and the retrieval results and the seasonal variations are presented in Section 3 with comparison to previous works, followed by a discussion in Section 4.

## 2. METHODOLOGY

### 2.1 Stellar Occultation Observations

Stellar occultation data used in this work are obtained using the Cassini/UVIS FUV spectrograph (Esposito et al. 2004). It has a spectral range of 1115-1912 Å with a resolution of ~1.5 Å, which covers the spectral extinction features of a number of small hydrocarbon and nitrile species, as well as haze particles (Figure 1). The initial data calibration and reduction are described in Chapters 9 and 10 in the *Cassini/UVIS Users Guide* (NASA-PDS 2017). The observed photon count spectra are firstly flat-fielded to remove the different detector sensitivities and then integrated by every ten observations to improve the statistical accuracy. Using the navigation data from the SPICE (NASA NAIF, Acton 1996), geometry information is derived, including the impact parameter (IP) and its location, as well as the pointing motion. Eighteen occultation observations are identified to have decent quality without significant data gaps (Figure 2 and Table 1). Their time coverage is longer than a season on Titan and crosses the spring equinox in 2009 (Figure 2). Vertical resolutions of these stellar occultations are the difference of the average IPs between two consecutive binned observations (Table 1), which vary from one flyby to another as a result of different spacecraft velocities during flybys.

### 2.2 Instrument Pointing Stability

A detailed description of the spacecraft's pointing motion and its influence on the observed spectrum are presented in Fan et al. (2019). As a compromise between keeping the spacecraft stable and saving fuels, the attitude control system of Cassini is designed to react only when the offset of the instrument pointing, referenced to the UVIS FUV principal axis, touches a deadband of 0.5 mrad during stellar occultations (Pilinski & Lee 2009), resulting in the motion of the target UV star in the instrument field-of-view (FOV). Due to the instrument's internal scattering, the star motion in FOV could cause non-linear shifts of characteristic spectral lines (e.g., the $C_2H_2$ line near 1520 Å) and/or photon scattering to extinction-saturated spectral regions (e.g., <1400 Å at small IPs, Fan et al. 2019). This influence prevented the data reduction of most of the stellar occultation observations, which requires a forward model that handles the pointing motion.

### 2.3 Chemical Species Cross Sections

UV extinction cross sections of the chemical species used in the retrieval are obtained from the *MPI-Mainz UV/VIS Spectral Atlas of Gaseous Molecules of Atmospheric Interest* (Keller-Rudek et al. 2013). It contains data from laboratory measurements. The temperature and corresponding wavelength range of each measurement are summarized in Table 2, together with their individual references, and the shapes of the cross sections are shown in Figure 1. As Titan's upper atmosphere is usually as cold as 150K, the cross-section measurements are selected to have temperature as low as possible, although for some species room-temperature measurements are the only available ones. The temperature difference may introduce ~20% uncertainties to the retrieved LOS abundances, but they do not influence the analyses of their seasonal changes, as the possible biases are systematic and consistent in all observations.

Eight hydrocarbon and nitrile species ($CH_4$, $C_2H_2$, $C_2H_4$, $C_2H_6$, $C_4H_2$, $C_6H_6$, HCN, $HC_3N$) and haze particles, which have extinction features in FUV (Figure 1), are considered in the retrieval. $C_2N_2$ and $C_6N_2$ are not considered, as the tests that included them did not show strong evidence of their existence (Fan et al. 2019). The haze particles are all assumed to be 12.5 nm spheres, same as that in previous works (Liang et al. 2007, Koskinen et al. 2011, Fan et al. 2019), and to have the same optical properties as their laboratory analogs, "tholins" (Khare et al. 1984).

## 2.4 LOS Abundance Retrieval

To address the instrument pointing motion, an instrument simulator (Shemansky et al. 2005, Shemansky & Liu 2012) is included in our forward model. The simulator takes in high-resolution UV spectrum and target star location in FOV to compute the corresponding photon count observations using line-by-line internal scattering response functions, which were measured before the launch of the spacecraft. The high-resolution UV spectrum is also wider (912-1950 Å) than that of the instrument due to the requirement of internal scattering computation.

At each iteration, the forward model first computes an extinction spectrum using proposed line-of-sight (LOS) abundances of species and their extinction cross sections. Then, the extinction spectrum is multiplied by a high-resolution spectrum of a typical UV star, normalized by the unattenuated target star spectrum ($I_0$) measured at IP>1500 km where the atmospheric extinction is negligible. Finally, the instrument simulator uses geometry information derived from navigation data to simulate a photon count spectrum, which is then compared with observations.

The MCMC approach (Mackay 2003) is used as a parameter-searching tool, which is implemented using a Python package *emcee* (Foreman-Mackey et al. 2013). The algorithm searches the posterior probability density functions (PDFs) of parameters, which are species LOS abundances in this case, with the capability to test the extent that each parameter can be constrained. As shown in Fan et al. (2019), some PDFs are flat or asymmetric with an upper limit only when the observation limit is reached because of too low or too high LOS abundances. The cost function in MCMC is defined as follows:

$$\ln(\mathrm{p}) = -\frac{1}{2}\sum_i \left[\frac{\left(I_{Obs_i} - I_{MCMC_i}\right)^2}{\sigma_i^2 + 0.1} + \ln(\sigma_i^2 + 0.1)\right] \qquad (1)$$

where p is the posterior probability; i is the wavelength index; $I_{Obs}$ and $I_{MCMC}$ are the observed and simulated photon count spectrum; $\sigma_i$ is the standard deviation at the i-th wavelength, which is assumed to be the square root of $I_{MCMCi}$. A softening factor of 0.1 is added to the variance to avoid dividing by zero when the spectrum is saturated. The MCMC starts with 100 chains and flat *a priori* in log space. In most of the cases, the equilibrium can be reached in the first 1000 steps, but sometimes it requires 4000-5000 steps. The final results used for seasonal variation analyses are obtained by running the MCMC for 1000 more steps after equilibrium.

By combining the forward model and the parameter-searching tool, PDFs of the LOS abundances of the species during all eighteen flybys are obtained. Each PDF is then fitted by three types of functions (Gaussian, sigmoid, and constant), and categorized by comparing the three fitting residuals. LOS abundances with Gaussian-like PDFs are defined as well-constrained. The values and uncertainties of these LOS abundances are therefore obtained. More details and validation of this retrieval method can be found in Fan et al. (2019).

### 2.5 Local Density Conversion

Vertical profiles of species local densities are derived using their corresponding LOS abundances. As not all species can be well-constrained in a large IP range, only species that have Gaussian-like LOS abundance PDFs at more than five IPs are converted to local densities. Firstly, a linear interpolation in log-space is applied to species LOS abundances and corresponding uncertainties within the region between the largest and smallest IPs where the LOS abundance is well-constrained. Secondly, the LOS abundance is linearly extrapolated, also in log space, above the well-constrained region until 2500 km, as the abundances above the upper IP boundary of the retrieval (1200 km) are not negligible. A 1000-step Bootstrap Monte Carlo (BSMC) method is then used for sampling the LOS species abundances from the PDFs, and finally the Abel inverse transform is applied with an assumption of spherical symmetry. Following this process, PDFs of local densities are derived. The resulting PDFs of local densities are also fitted and categorized the same way as those of LOS abundances using three types of functions (Gaussian, sigmoid, and constant).

## 3. RESULTS

### 3.1 Example Retrieval Result

A detailed step-by-step validation of the retrieval process is presented in Fan et al. (2019). Here, a typical spectrum used in the retrieval (Figure 3) is shown for the purpose of illustration, which is obtained at ~756 km during flyby T52 on 2009/093. The forward model agrees well with the observation, with residual symmetrically distributed around zero. The residual mostly comes from the offset of featured species spectral lines, which is likely due to the temperature and pressure differences between Titan’s atmosphere and the laboratory environments where the cross sections were measured.

Figure 4 shows the probability density functions (PDFs) of the retrieved parameters, the LOS abundances of the nine species, using the observed spectrum shown in Figure 3. As described in Section 2.4, three types of functions (Gaussian, sigmoid, and constant) are fitted to these PDFs to quantify the extent to which they are constrained, and these PDFs are then categorized by comparing the fitting residuals. In this particular case, most of the species LOS abundances are well-constrained with Gaussian-like PDFs, except for $C_2H_6$ whose upper limit can only be obtained due to its similar extinction cross section shape with that of $CH_4$ (Figure 1), and for haze which is due to its small spectral contribution at this IP.

## 3.2 Seasonal Variation of Chemical Species

Eighteen occultation observations with decent data quality are considered in this analysis (Figure 2 and Table 1). They cover a time range from 2006 (T13, $L_S$=318°) to 2014 (T105, $L_S$=60°). There is a gap immediately after the equinox between 2010 and 2014 when UVIS stellar occultations were not scheduled during Titan flybys of Cassini. According to the latitude distribution of flybys (Figure 2), from 42.8°S (T56) to 55.5°N (T47e), observations are grouped for the investigation of seasonal variations: Group 1 in the equatorial region between 20°S and 20°N, including T13, T23, T41i, T47i, T48, and T58; Group 2 south of 20°S, including T21, T35, T41e, T56, T101e, T103, and T105; Group 3 north of 20°N, including T40, T47e, T52, T53, and T101i. LOS abundances are used for the comparison instead of local number densities, as the former ones are the quantities retrieved directly from observations, which therefore have smaller uncertainties and larger well-constrained ranges.

The profiles of the eight chemical species are shown in Figure 5 and 6, and the values are presented in the Supplementary Materials of this paper. Seasonal variations of the chemical species indicate the changes of physical and chemical processes. The abundance of $CH_4$, the second most abundant chemical component in Titan's atmosphere with a mixing ratio of a few percent (Shemansky et al. 2005, Koskinen et al. 2011, Kammer et al. 2013), is an indicator of the physical properties and dynamics of Titan's upper atmosphere when measurements of the most abundant species $N_2$ are not available. Yung et al. (1984) proposed the major schemes of the productions of hydrocarbons and nitriles, which are reevaluated and updated using Cassini observations (Ågren et al. 2009, Galand et al. 2010, Lavvas et al. 2011, Loison et al. 2019, Vuitton et al. 2019). Abundances of small hydrocarbons ($C_2H_2$, $C_2H_4$, $C_2H_6$, $C_4H_2$, $C_6H_6$) are tracers of $CH_4$ photochemistry. They are immediate chemical products of the organic reactions, where some of them also serve as catalysts. The production of HCN from the two dominant species ($N_2$ and $CH_4$) is dominated by solar photons but influenced by energetic electrons, so the abundance of HCN is also an indicator of Titan's space environment besides solar radiation. The abundance of $HC_3N$ is controlled by those of $C_2H_2$ and HCN, as they are the reactants in the major production reaction. Therefore, the investigation of seasonal variations below mainly focuses on two $C_2$ species ($C_2H_2$ and $C_2H_4$) and HCN for two reasons: (1) all hydrocarbons have similar behaviors due to their formation pathways, among which $C_2H_2$ and $C_2H_4$ are the best constrained (Figure 5 and 6); and (2) the seasonal variation of $HC_3N$ can be well explained by the changes of $C_2H_2$ and HCN. The constraint on $C_2H_6$ is poor (Figure 5d, 5h, and 5l) due to its overlapping spectral feature with $CH_4$ (Figure 1, Fan et al. 2019), so it is excluded from the discussion.

In the equatorial region between 20°S and 20°N (Group 1, Figure 5a-5d and 6a-6d), results obtained during the six flybys do not show noticeable differences. The profiles of $CH_4$ LOS abundances are similar among observations in the well-constrained range 500-1200 km, and follow exponential distributions with a scale height of ~64 km (Figure 5a). Profiles of $C_2H_2$ and $C_2H_4$ also follow similar exponential decay at 700-1000 km, while some oscillations exist in the IP range of 400-700 km (Figure 5b and 5c). This likely results from temperature perturbations, which are suggested to be as large as 20-30 K with a vertical length scale of ~50 km (Koskinen et al. 2011). Ratios of $C_2H_2$ and $C_2H_4$ abundances to that of $CH_4$ peak at 800 km with values of ~5 percent due to methane photolysis, and decrease below when the UV absorption gradually reaches saturation (Figure 7a and 7b). Profiles of HCN show similar features, which are well constrained below 1000km with peaks near ~750km (Figure 6c). Altitude of the peaks is a combined effect of solar photon absorption and energetic electron collision of $N_2$, whose rates become larger with increasing air density when above ~750 km, and decrease due to the saturation at lower altitudes. The similarity among profiles of all species during different flybys suggests that in the time range before the equinox (early-2007 to mid-2009), there is no significant change in the atmospheric characteristics near the equator.

The results obtained during the seven flybys south of 20°S (Group 2) show significant seasonal variations (Figure 5e-5h, and 6e-6h). The $CH_4$ LOS abundance remains the same before the equinox, but it reduces by more than half at the end in the upper atmosphere (Figure 5e). It is ~$10^{17}$ $cm^{-2}$ at ~900 km during all four flybys before equinox (T21, T35, T41e, and T56), and drops to ~$4*10^{16}$ $cm^{-2}$ during the last two (T103 and T105). This seasonal change is likely due to the decrease of Titan's atmospheric temperature, as the Saturn system kept moving away from the Sun during this period with a perihelion in mid-2003 and an aphelion in early-2018. The smaller scale height of Titan's atmosphere results in lower air density at given high altitudes. Moreover, as $CH_4$ in Titan's atmosphere is likely replenished by surface/subsurface processes (Yung et al. 1984), its abundance in the upper atmosphere is highly influenced by the condensation in the troposphere near 8 km (Atreya et al. 2006). Given the fact that saturation vapor pressure is sensitive to temperature, a small decrease of temperature in the $CH_4$ condensation region could result in a large abundance drop in the upper atmosphere. This scenario is supported by the $CH_4$ LOS abundance profile of T101e (2014/137), which is similar to those before the equinox above 1000 km, and gradually converges to the profiles of T103 (2014/201) and T105 (2014/265) at smaller IPs. As T101e is only ~60 days earlier than T103 and they are within 1° in latitude (Figure 2 and Table 1), the difference in the upper atmosphere above ~800 km shows evidence of a short-time-scale transitional stage of this mechanism. Also, as this region is above methane's homopause (~800 km, Li et al. 2014), the high-altitude difference is not likely due to atmospheric dynamics. LOS abundances of $C_2H_2$ and $C_2H_4$ show the same behavior as that of $CH_4$ (Figure 5f and 5g), and the ratios of them to that of $CH_4$ do not have significant difference (Figure 7c and 7d), which suggest that there is not much difference in photochemistry, and the seasonal variation mostly results from the change of atmospheric temperature. Similarly, the LOS abundances of HCN also have significant decreases after the equinox (Figure 6g), despite their large uncertainties. This is likely due to the decrease of atmospheric temperature as well, which results in a smaller scale height of the atmosphere and therefore a smaller abundance of $N_2$ at a given altitude in the upper atmosphere.

LOS profiles of Group 3 in the region north of 20°N present potential evidence of seasonal variation of dynamics. The $CH_4$ LOS abundance does not show much difference before and after the equinox (Figure 5i), while the decrease of $C_2H_2$ and $C_2H_4$ abundances are more than a factor of ~2 at most of the IPs (Figure 5j and 5k), which is also shown more clearly by the abundance ratios to $CH_4$ (Figure 7e and 7f). As solar insolation increases in the northern hemisphere after the equinox, which results in more intensive photochemistry, the decreases of $C_2H_2$ and $C_2H_4$ abundances are likely due to the change of atmospheric circulation. Circulation in Titan's middle and lower atmosphere has been well simulated and understood in the last decade (e.g., Newman et al. 2011, Lebonnois et al. 2012, Lora et al. 2015). Due to Titan's slow rotation (~16 days), one Hadley cell extends pole to pole with ascending motion in the summer hemisphere and descending in the winter, while two cells appear during transition near the equinox. However, the altitude of the top of the Hadley cell is poorly constrained, and is suggested to be above the top boundary of most of the numerical simulations (400-500 km) and to be at least 600 km (Teanby et al. 2012). The smaller mixing ratios of $C_2H_2$ and $C_2H_4$ in the summer hemisphere indicate that the upwelling branch of the Hadley cell can likely extend to Titan's homopause (~800 km) and to latitude as low as ~45°. It transports the air with smaller mixing ratios from lower altitude upwards in the summer hemisphere without changing the abundance of $CH_4$. The HCN abundance obtained in this region also decreases significantly after the equinox (Figure 6k), which is consistent to that in Group 2, and supports the scenario of temperature decrease in Titan's atmosphere.

### 3.3 Detached Haze Layer

Altitude of the detached haze layer is identified as the peak of the haze LOS abundance profiles (Figure 8a-8c). The haze particles are represented by 12.5 nm spheres in the retrieval, as the spectral slope of their UV extinction cross sections are not sensitive to the particle radii when haze is the major factor causing the change of total UV extinction (Koskinen et al. 2011). A decrease of the detached haze layer altitude with time is found before the equinox, but afterwards the existence of such peaks is not clear (Figure 8). The detached haze layer is either absent or at an altitude below the lower boundary (400 km) of the retrieval range. The altitude of the detached haze layer is also compared with observations made by the visible camera, the Imaging Science Subsystem (ISS) onboard Cassini (West et al. 2011), which shows good agreement between results of the two instruments (Figure 8d), expect for a small disagreement of ~20 km immediately before the equinox. This serves as a cross validation of the two independent observations.

The detached haze layer and the decrease of its altitude before the equinox has been reported in many previous works (e.g., West et al. 2011, Koskinen et al. 2011, West et al. 2018, Seignovert et al. 2021), but the mechanism of its origin and seasonal variation is still an open question. Several explanations have been proposed to address this issue, e.g., transition of the fractal dimension of haze particles (Lavvas et al. 2009), atmospheric dynamics and seasonal change of the Hadley cell (Lebonnois et al. 2012), and gravity waves (Koskinen et al. 2011), but approval or negation of them requires further investigations. Results from this work provide observational constraints.

## 3.4 Vertical Local Density Profiles

In Section 3.2, seasonal variations in Titan's upper atmosphere are investigated using LOS abundances of chemical species due to their better constraints than local densities. However, vertical profiles of these local densities (Figure 9 and 10) are necessary for future comparison with photochemical models and/or GCMs. The data of these profiles are presented in the Supplementary Materials of this paper. As the local density at a given altitude depends on all LOS abundances at IPs that are equal and above this altitude, they usually have larger relative uncertainties and smaller well-constrained regions than corresponding LOS abundances. Although these profiles are relatively poorly constrained, similar seasonal variations can still be identified for most of the species ($CH_4$, $C_2H_2$, $C_2H_4$, $C_4H_2$, $HC_3N$), while for the others with large LOS abundance uncertainties ($C_2H_6$, $C_6H_6$, HCN), the vertical profile conversion does not result in good constraints.

## 3.5 Comparison with Previous Retrievals

Results obtained during two flybys, which are also considered in previous works using UVIS observations, are selected for the comparison of retrieval processes: T41i (2008/054) with Koskinen et al. (2011) and T52 (2009/093) with Fan et al. (2019). LOS abundances of four species, $CH_4$, $C_2H_2$, $C_2H_4$ and HCN, are included in the comparison (Figure 11). Comparison of T41i shows good agreement between the two retrieval methods (Figure 11a-11d) within the well-constrained regions, except for HCN at >800 km where its LOS abundance is relatively poorly constrained. Koskinen et al. (2011) applied the Levenberg–Marquardt (LM) algorithm onto measured UV optical depths with species extinction cross sections. As computing the optical depths, which includes dividing one spectrum by another, requires the pointing motion to be small during the entire flyby, this method is limited to stable flybys, e.g., T41i and T53 as included in Koskinen et al. (2011). The LM algorithm also assumes symmetric posteriors of parameters, LOS abundances in this case. However, in regions where the spectral information of a species is not sufficient, usually at small or large IPs, the symmetry assumption does not stand. This likely leads to the main difference between the two retrieval results: <600 km for $CH_4$ (Figure 11a), >1000 km for $C_2H_2$ and $C_2H_4$ (Figure 11b and 11c), and >800 km for HCN (Figure 11d). The comparison of T52 with Fan et al. (2019) shows exceptionally good agreement (Figure 11e-11h), as the updated UV extinction cross sections and the exclusion of two testing species ($C_2N_2$ and $C_6N_2$) are the only difference between these two retrievals. Small differences in the cross sections and the consideration of the two species do not reflect much difference in the retrieved LOS abundances of major species.

## 3.6 Comparison with Other Observations

Vertical profiles of the same four species ($CH_4$, $C_2H_2$, $C_2H_4$ and HCN) are also compared with the results obtained by two other instruments onboard the Cassini spacecraft, at high altitudes (>1000 km) with the Ion Neutral Mass Spectrometer (INMS, Waite et al. 2004,

Cui et al. 2009) which includes the effect of instrument recalibration (Teolis et al. 2015), and at low altitudes (<600 km) with the Composite InfraRed Spectrometer (CIRS, Flasar et al. 2004, Mathé et al. 2020). Results during flybys before the equinox in Group 3 (north of 20°N) are selected for the comparison (Figure 12), as the INMS data (Cui et al. 2009) are mostly in the northern mid- and high-latitude regions and within this time range. The comparison shows good agreement among the results by the three instruments, despite some small differences. The INMS results of $CH_4$ local densities are slightly larger than those of UVIS (Figure 12a), which is likely a result of the earlier season of INMS observations when the solar insolation is stronger due to smaller solar distance. The CIRS results suggest stronger oscillations and seasonal variations, likely due to atmospheric dynamics in the middle atmosphere.

## 4. DISCUSSION

Stellar occultation observations obtained by Cassini/UVIS have significantly increased our understanding of physical and chemical processes in Titan's upper atmosphere. During the last decades, selected flybys with stable pointing motions have been analyzed (TB, T21, T35, T41i, T41e, T116, Shemansky et al., 2005, Koskinen et al., 2011, Capalbo et al., 2016, Yelle et al. 2021). Fan et al. (2019) proposed a new method to address the pointing issue, and successfully analyzed the observation during a flyby with large pointing motion (T52). The new method is applied to eighteen observations with decent data quality in this work and with updated cross section measurements for chemical species. The flybys overlapping with previous works are reanalyzed in this work with our updated retrieval scheme. Comparison of the results derived from two overlapped flybys (T41i with Koskinen et al. 2011, T52 with Fan et al. 2019) shows good agreements (Figure 11). The new method gives results similar to the traditional approach in retrieving species abundances from pointing-stable occultation observations, but it is much more capable when the pointing motion is large.

In the analysis of seasonal variations, the species abundances at similar latitude before and after the equinox show significant and consistent differences, which are indications of the change of Titan's atmospheric structure and dynamics. However, some features seem to be transient in time and are different among species. During four flybys, T23 (2007/013), T47i (2008/324), T56 (2009/157), and T58 (2009/189), the LOS abundances of $C_2H_4$ and $C_6H_6$ are significantly higher at large IPs (>900 km) than those of other flybys, while similar features are only noticeable during some flybys for $C_2H_2$ and $HC_3N$ (Figure 5 and 6). As the retrieval is independently applied to each IP, the deviation of LOS abundances from normal trends at a few consecutive IPs suggests structures in Titan's atmosphere, which may be due to some transient mechanisms, e.g., solar activity or rapid change of transport. Future photochemical or dynamic modeling should address the variability of the hydrocarbons due to transient forcing.

As the eighteen flybys span a long period of time (~9 years), which covers different seasons of Titan and crosses the spring equinox, investigation of the seasonal change in Titan's upper atmosphere becomes possible. Several photochemical models have been developed in simulating the profiles of chemical species (e.g., Yung et al. 1984, Li et al. 2014 and 2015, Willacy et al. 2016, Loison et al. 2019, Vuitton et al. 2019), but results from stellar occultation observations during only five flybys, TB, T21, T41i, T41e, and T53 (Shemansky et al. 2005, Koskinen et al. 2011, Capalbo et al. 2016), and dayglow observations during three flybys, T55, T81, and T94 (Stevens et al. 2015) have been used so far to constrain the chemistry in Titan's upper atmosphere. Results presented in this work greatly expand the number of observations and provide constraints for investigations of seasonal change of photochemistry and dynamics. Besides, general circulation models have also been developed to study seasonal changes in Titan's lower and middle atmosphere (e.g., Newman et al. 2011, Lebonnois et al. 2012, Lora et al. 2015), but their altitude upper boundaries are mostly at 400-500 km, because the approximations of the governing equations become invalid if applied to the upper atmosphere. Müller-Wodarg et al. (2000, 2003) developed a GCM to predict the dynamics in Titan's upper atmosphere, but it is not sufficiently constrained by observations. Results from the present work can likely address this issue. Development and improvement of photochemical models and GCMs are anticipated.

## 5. SUMMARY

Eighteen stellar occultation observations obtained by Cassini/UVIS over one fourth of Titan's year are analyzed in this work. Abundances of eight chemical species ($CH_4$, $C_2H_2$, $C_2H_4$, $C_2H_6$, $C_4H_2$, $C_6H_6$, HCN, $HC_3N$) and haze particles are constrained in Titan's upper atmosphere between 400 and 1200 km. Seasonal changes of vertical profiles of these species suggest variations in atmospheric structure and dynamics. The detached haze layer moves downward before the spring equinox, and is not identified afterwards. This work provides observational constraints of Titan's upper atmosphere, and suggests simulations to further understand chemistry and dynamics.

## ACKNOWLEDGMENTS

This research was supported in part by the Cassini/UVIS program via NASA Grant JPL.1459109 to the California Institute of Technology and was partially supported by funding from NASA's Astrobiology Institute's proposal "Habitability of Hydrocarbon Worlds: Titan and Beyond" (PI R.M. Lopes). All the data and tools in this work are publicly available. Cassini/UVIS data are available on NASA PDS (pds.nasa.gov). The python package emcee is available online (dfm.io/emcee/current). We thank Tommi T. Koskinen for sharing Cassini/UVIS results, Jun Cui for sharing Cassini/INMS results and comments, Sandrine Vinatier and Christophe Mathé for sharing Cassini/CIRS results, and Karen Willacy for sharing photochemical model results.

## TABLES

**Table 1 Stellar Occultations**

| Flyby | Date [YYYY/DOY] | $L_S$ [°] | Star | Latitude [°] | Longitude [°] | Resolution [km] |
|---|---|---|---|---|---|---|
| T13 | 2006/120 | 318 | β Ori | -16.72 | 312.04 | 52.99 |
| T21 | 2006/346 | 326 | α Eri | -35.5 | 118.47 | 37.99 |
| T23 | 2007/013 | 327 | η UMa | -4.6 | 231.82 | 23.82 |
| T35 | 2007/243 | 335 | σ Sgr | -33.25 | 327.45 | 58.25 |
| T40 | 2008/005 | 340 | α Lyr | 53.26 | 33.22 | 30.51 |
| T41i | 2008/054 | 341 | ε CMa | -7.49 | 332.43 | 4.42 |
| T41e | 2008/054 | 341 | ε CMa | -26.92 | 174.61 | 7.78 |
| T47i | 2008/324 | 351 | η UMa | 0.72 | 24.03 | 62.91 |
| T47e | 2008/324 | 351 | β CMa | 55.52 | 340.12 | 55.73 |
| T48 | 2008/340 | 351 | ε CMa | 18.64 | 320.21 | 65.41 |
| T52 | 2009/093 | 356 | α Eri | 37.12 | 312.8 | 25.06 |
| T53 | 2009/109 | 356 | α Eri | 39.02 | 296.02 | 16.48 |
| T56 | 2009/157 | 358 | η UMa | -42.76 | 80.25 | 16.32 |
| T58 | 2009/189 | 359 | η UMa | -11.4 | 31.44 | 41.4 |
| T101i | 2014/136 | 56 | η UMa | 45.89 | 23.91 | 18.81 |
| T101e | 2014/137 | 56 | η UMa | -31.5 | 138.06 | 7.58 |
| T103 | 2014/201 | 58 | α Eri | -30.85 | 247.5 | 6.88 |
| T105 | 2014/265 | 60 | η UMa | -25.5 | 133.58 | 28.73 |

Note: Latitude and longitude are computed at IP=500 km.

**Table 2 Extinction cross sections**

| Species | T [K] | Wavelengths [Å] | Reference |
|---|---|---|---|
| $CH_4$ | 298 | 912-1246 | Kameta et al. (2002) |
| | 295 | 1246-1530 | Chen & Wu (2004) |
| $C_2H_2$ | 298 | 912-1200 | Cooper et al. (1995) |
| | 150 | 1200-1900 | Wu et al. (2001) |
| | 195 | 1900-1950 | Smith et al. (1991) |
| $C_2H_4$ | 298 | 912-1050 | Holland et al. (1997) |
| | 298 | 1050-1750 | Lu et al. (2004) |
| | 295 | 1750-1950 | Orkin et al. (1997) |
| $C_2H_6$ | 298 | 912-1200 | Au et al. (1993) |
| | | 1600-1650 | |
| | 295 | 1200-1600 | Chen & Wu (2004) |
| HCN | 298 | 912-1469 | Nuth & Glicker (1982) |
| | 298 | 1469-1543 | Lee et al. (1980) |
| $C_4H_2$ | 173 | 1150-1690 | Ferradaz et al. (2009) |
| | 223 | 1690-1950 | Fahr & Nayak (1994) |
| $C_6H_6$ | 298 | 912-1153 | Rennie et al. (1998) |
| | 298 | 1153-1715 | Capalbo, et al. (2016) |
| | 215 | 1715-1950 | Capalbo et al. (2016) |
| $HC_3N$ | 298 | 912-1120 | Ferradaz et al. (2009) |
| | | 1620-1950 | |
| | 203 | 1120-1620 | Ferradaz et al. (2009) |

Note: The cross section of haze particles is computed assuming that they consist of 12.5 nm spheres and have chemical compositions same as “tholins” (Khare et al. 1984).

## FIGURES

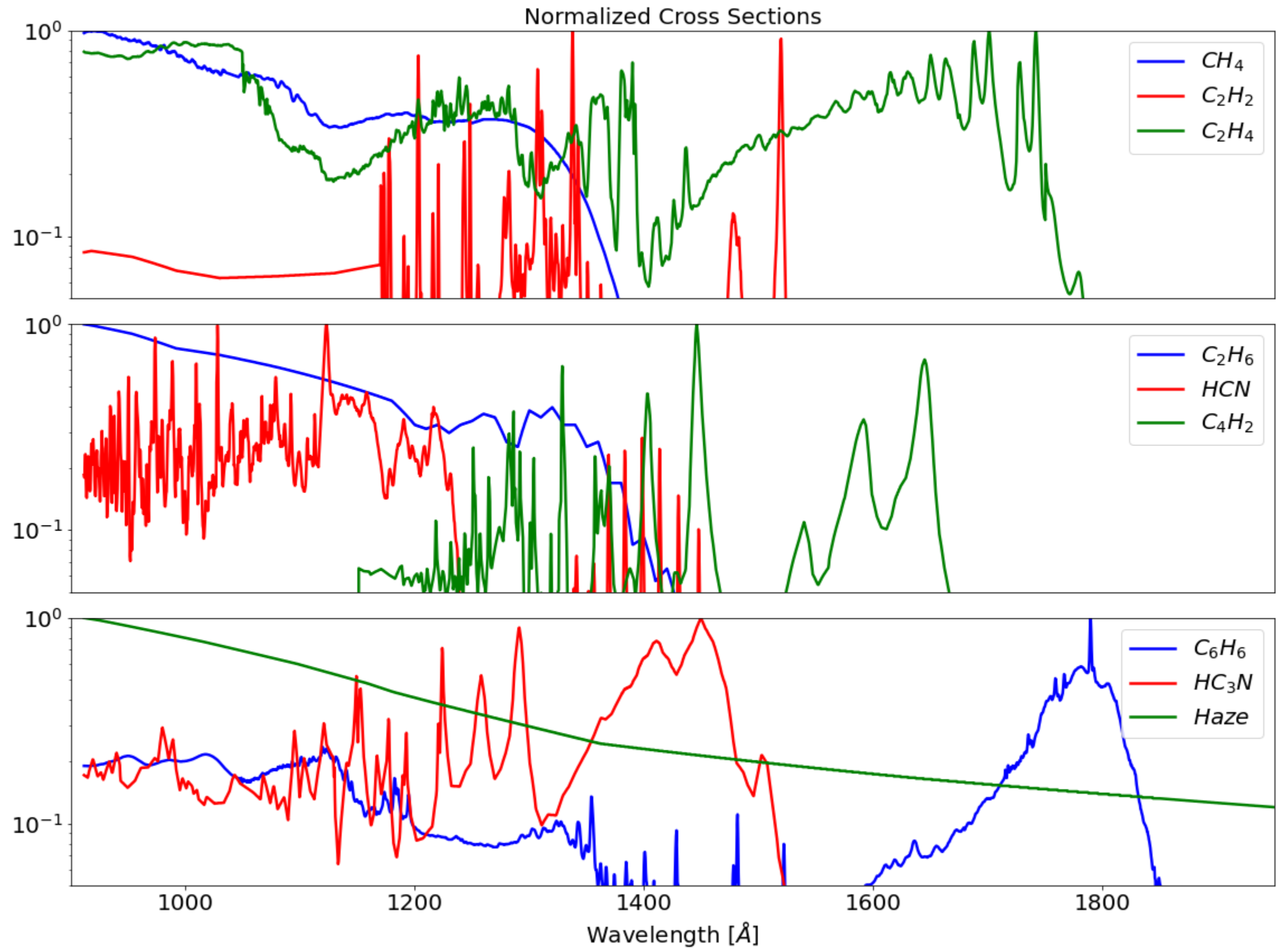

**Figure 1.** Normalized UV extinction cross sections of species considered in this work. Data of these cross sections are obtained from the *MPI-Mainz UV/VIS Spectral Atlas of Gaseous Molecules of Atmospheric Interest* (Keller-Rudek et al. 2013). The cross sections are normalized to have maximum values of unity.

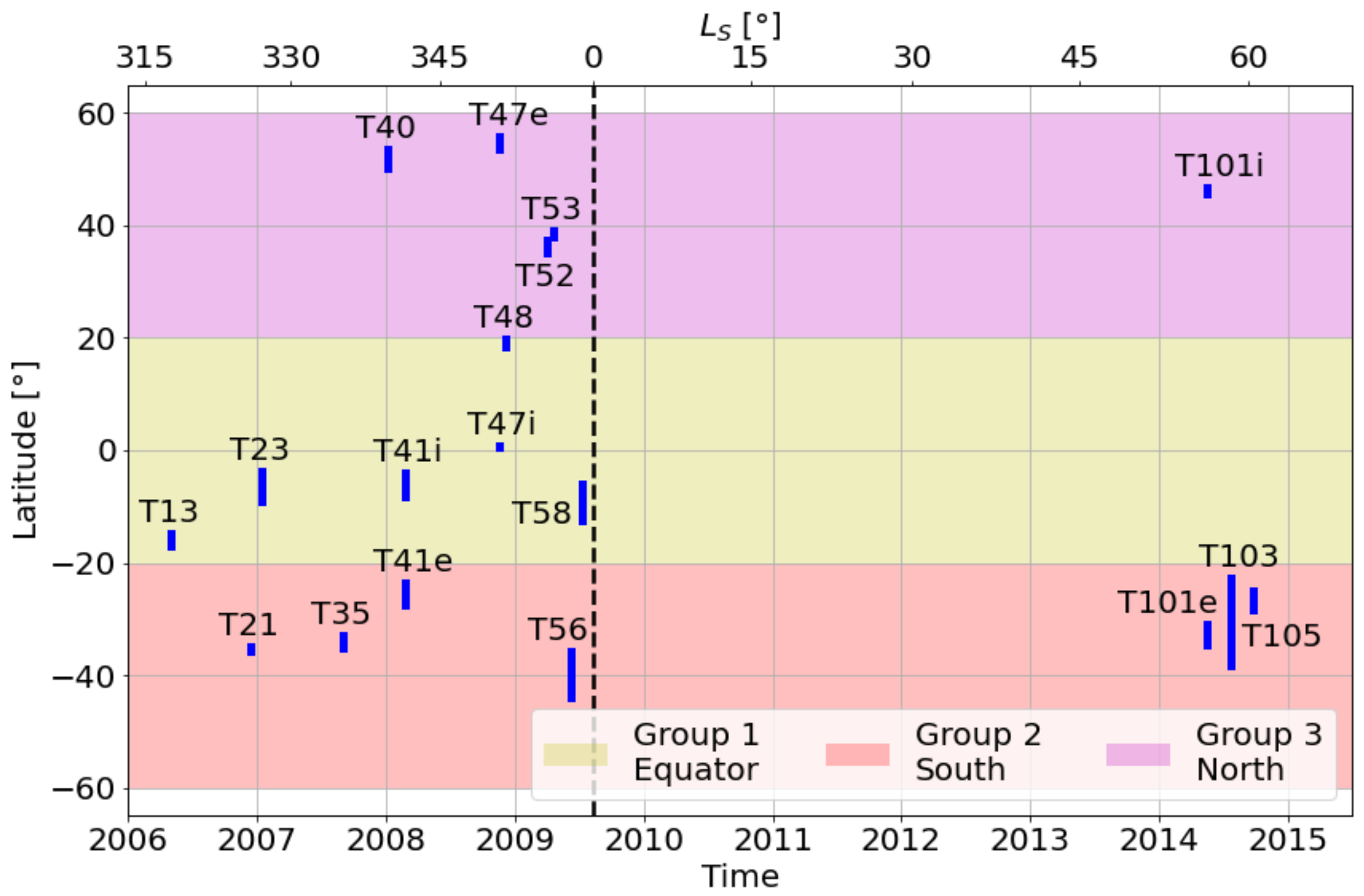

**Figure 2.** Time and location of stellar occultation observations. The latitudes range (blue lines) are computed when the impact parameter (IP) is between 400-1200 km during each flyby. Locations of the numbers on the horizontal axis denote day 1 of the corresponding years. The vertical black line denotes the time of the spring equinox. The three groups of flybys used for analysis of Titan's seasonal changes are denoted as colored shaded areas: Group 1 in the equatorial region (20°S to 20°N, yellow shaded area), Group 2 south of 20°S (red shaded area), and Group 3 north of 20°N (magenta shaded area).

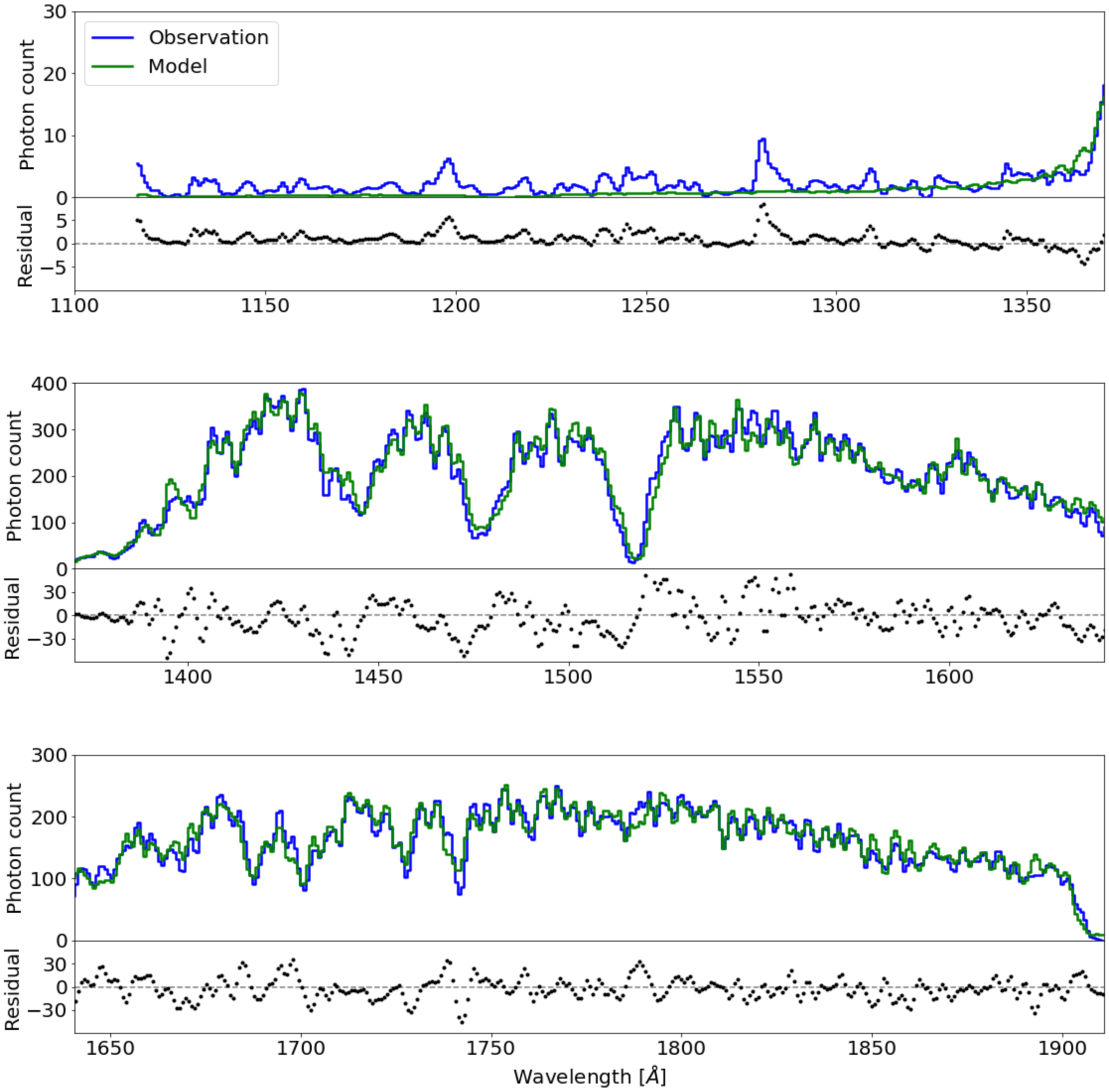

**Figure 3.** Spectra of photon counts and residuals of T52 at 756 km. The observation and synthetic spectrum are denoted as blue and green lines, respectively, and the residuals are denoted as black dots. The scales of the y-axes are not the same for the purpose of illustration.

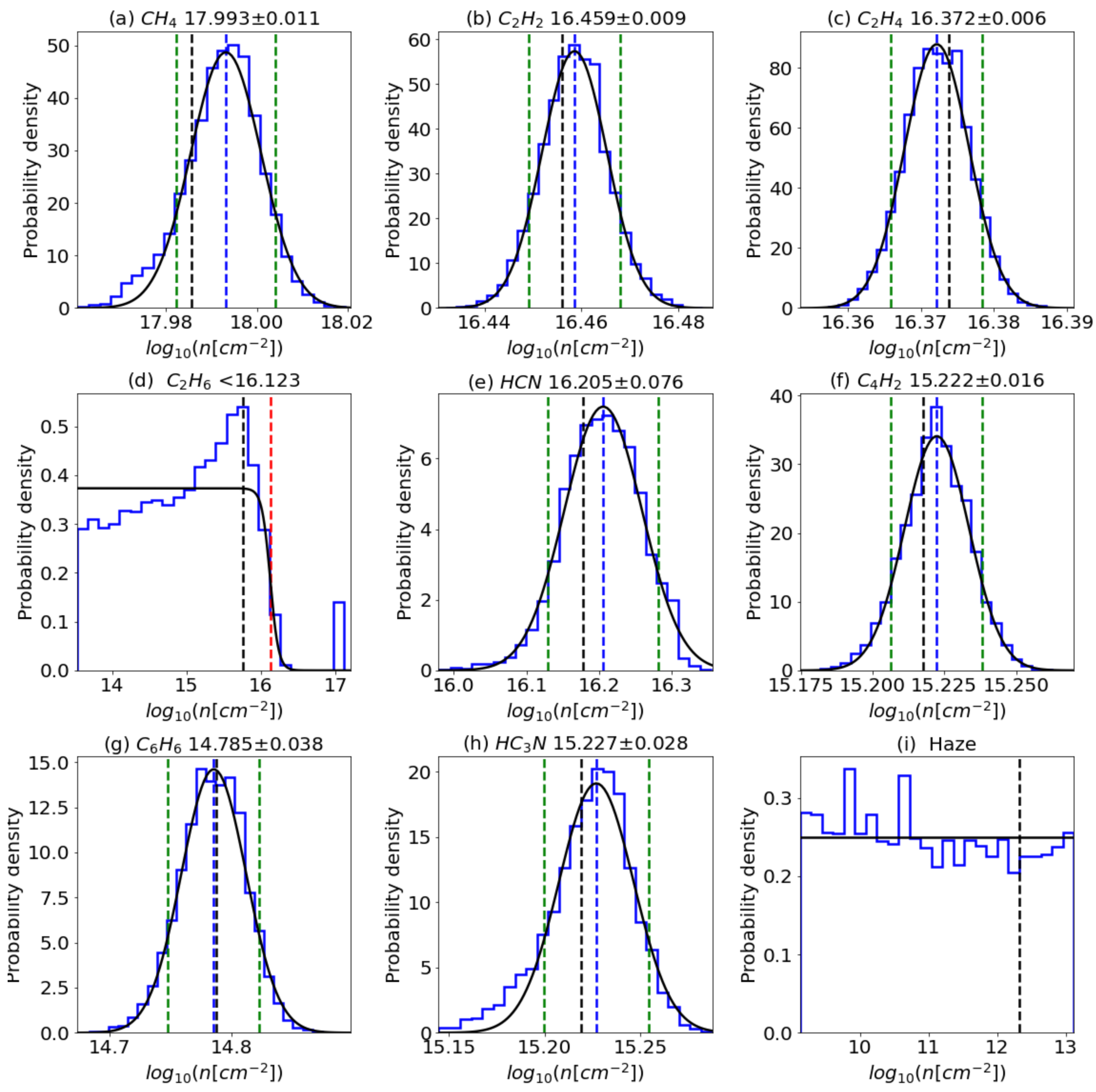

**Figure 4.** Probability density functions (blue solid lines) of the line of sight (LOS) abundances of (a) $CH_4$, (b) $C_2H_2$, (c) $C_2H_4$, (d) $C_2H_6$, (e) HCN, (f) $C_4H_2$, (g) $C_6H_6$, (h) $HC_3N$, and (i) haze, retrieved using the observations of T52 at 756 km. The corresponding spectra are shown in Figure 3. The best-fit function to each probability density function (Gaussian, sigmoid, or constant) is shown as black solid lines. The median and 1σ confidence interval are denoted by blue and green dashed lines, respectively, for well-constrained species. The upper limits are denoted by red dashed lines when the upper limit can only be constrained. Black dashed lines denote the values of LOS abundances whose combination reaches the maximum likelihood in the retrieval. Values of the medians and uncertainties or those of the upper limits are shown in the title of each panel.

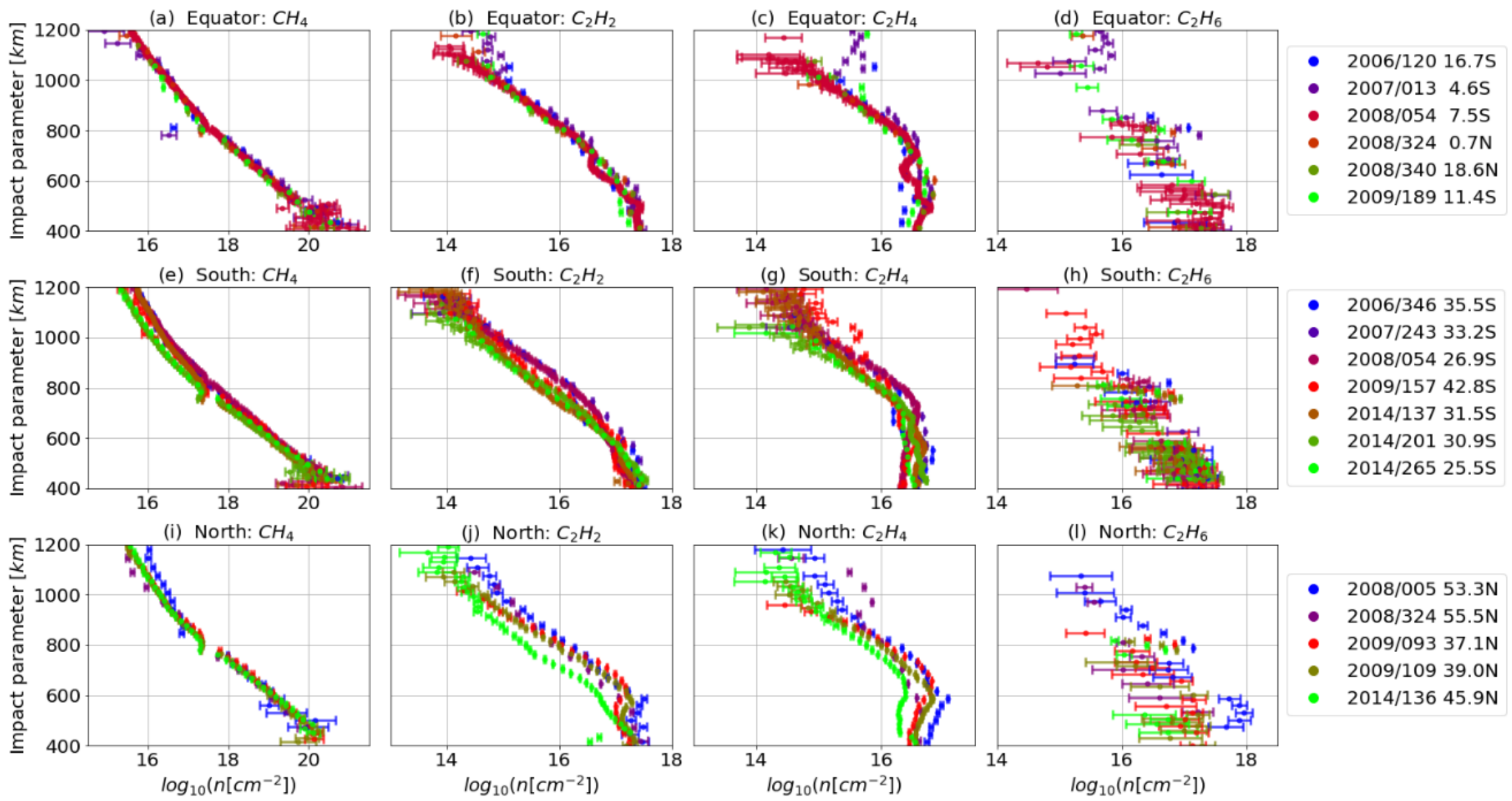

**Figure 5.** (a) LOS abundances of $CH_4$ retrieved from stellar occultation observations during Titan flybys in the equatorial region between 20°S and 20°N (Group 1). (b), (c), and (d), same as (a), but for $C_2H_2$, $C_2H_4$, and $C_2H_6$, respectively. Time and location of the flybys are denoted on the right of the four panels. (e)-(h), same as (a)-(d), respectively, but for flybys south of 20°S (Group 2). (i)-(l), same as (a)-(d), respectively, but for flybys north of 20°N (Group 3).

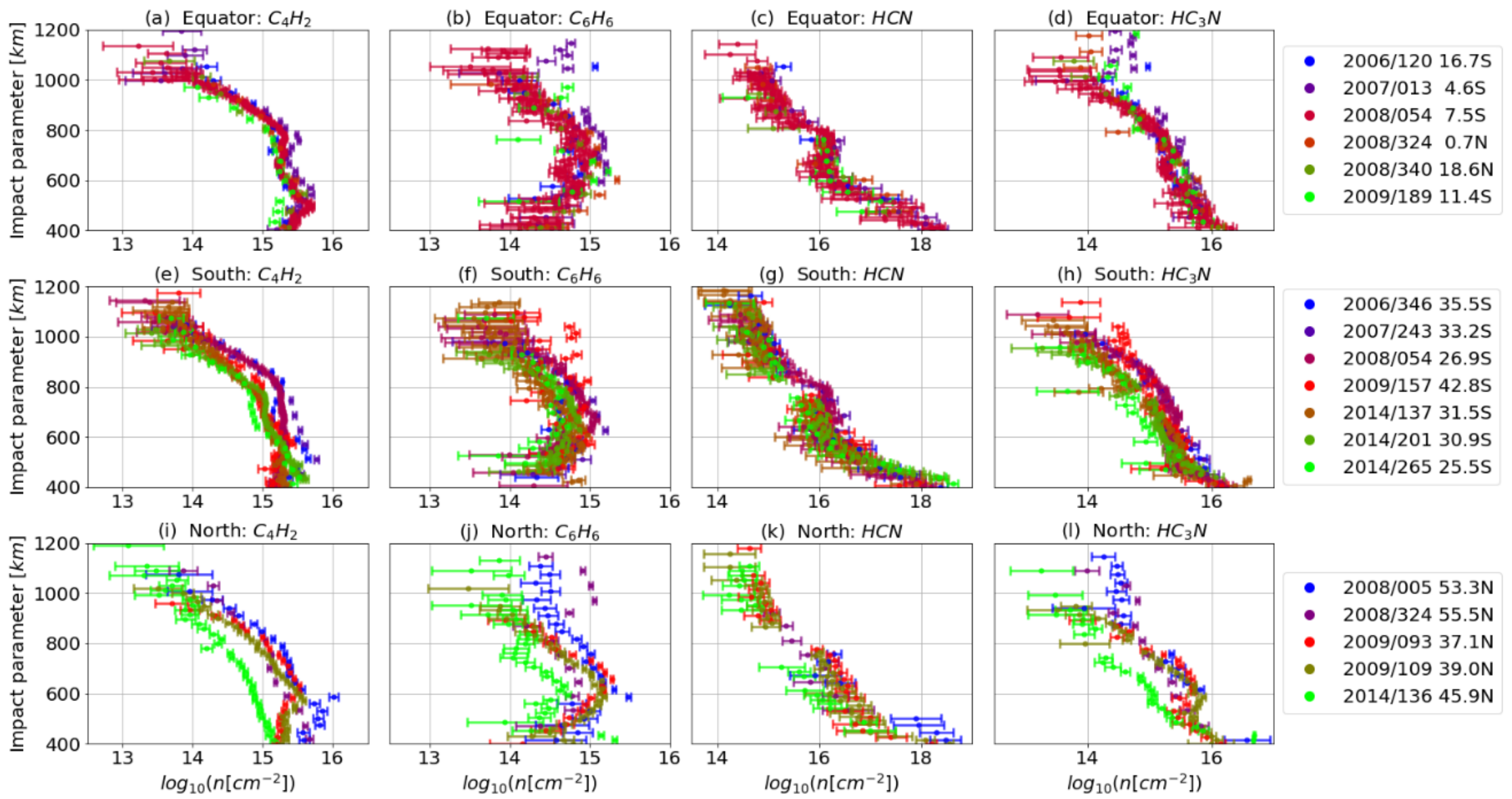

**Figure 6.** Same as Figure 5, but for (a, e, i) $C_4H_2$, (b, f, j) $C_6H_6$, (c, g, k), HCN, and (d, h, l) $HC_3N$.

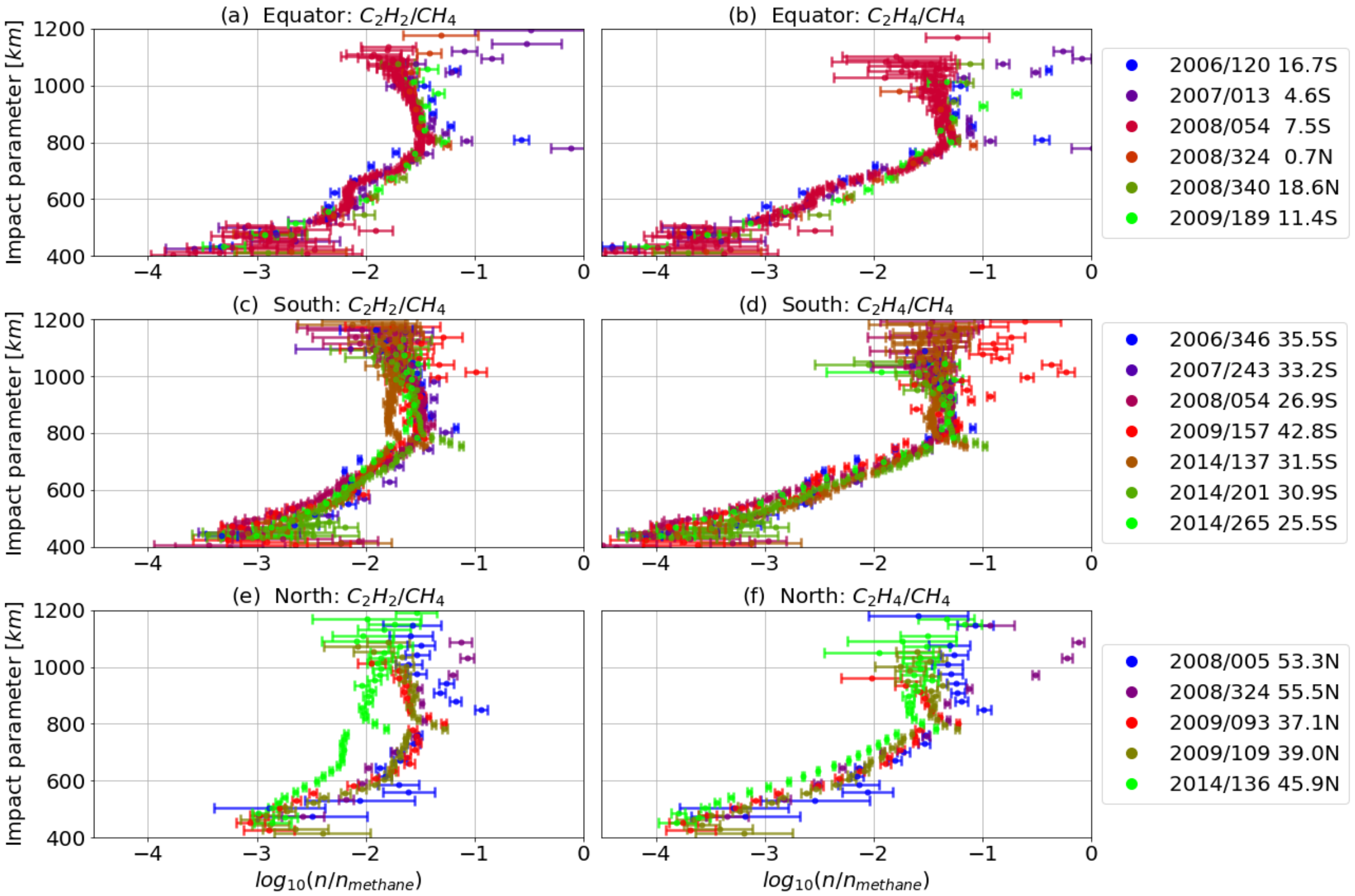

**Figure 7.** (a) Ratio of LOS abundances of $C_2H_2$ to $CH_4$ of flybys in the equatorial region between 20°S and 20°N (Group 1). (b) Same as (a), but for $C_2H_4$ to $CH_4$. Time and location of the flybys are denoted on the right of the two panels. (c) and (d), same as (a) and (b), respectively, but for flybys south of 20°S (Group 2). (e) and (f), same as (a) and (b), respectively, but for flybys north of 20°N (Group 3).

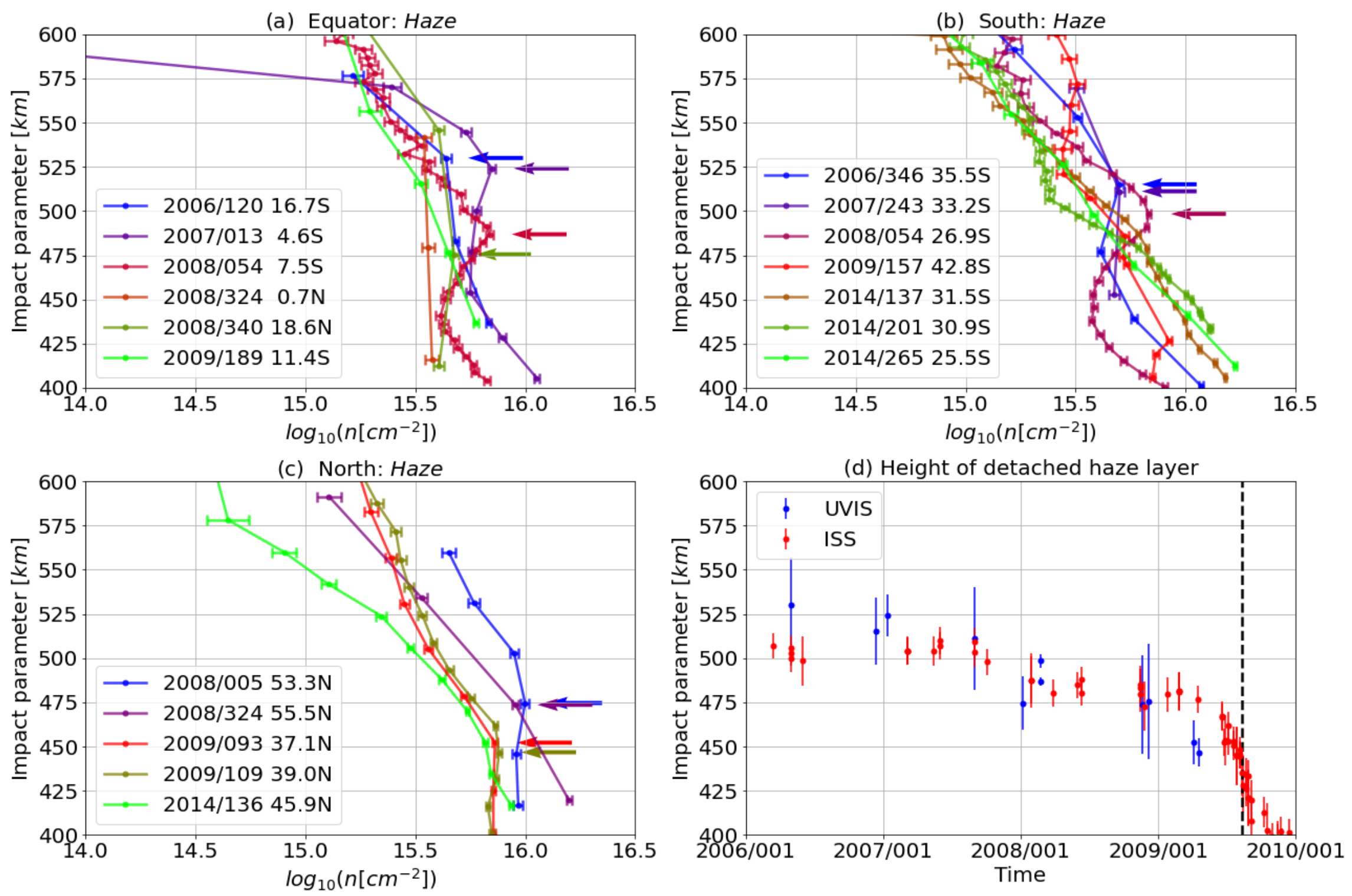

**Figure 8.** (a) LOS abundances of haze particles of flybys in the equatorial region between 20°S and 20°N (Group 1). The arrows denote the altitudes of the detached haze layers. Time and location of the flybys are denoted in the legend. (b) and (c), same as (a), but for flybys south of 20°S (Group 2) and north of 20°N (Group 3), respectively. T56 (2009/157) is excluded in determining the altitude of the detached haze layer due to a data gap at 426-470 km. (d) Altitudes of detached haze layers obtained using the UVIS observations (blue dots with error bars, this work), and the ISS (red dots with error bars, West et al. 2009). The time of the spring equinox is denoted as the black vertical dashed line. Vertical uncertainties of UVIS results are defined as half of the vertical resolution of the corresponding flyby (Table 1).

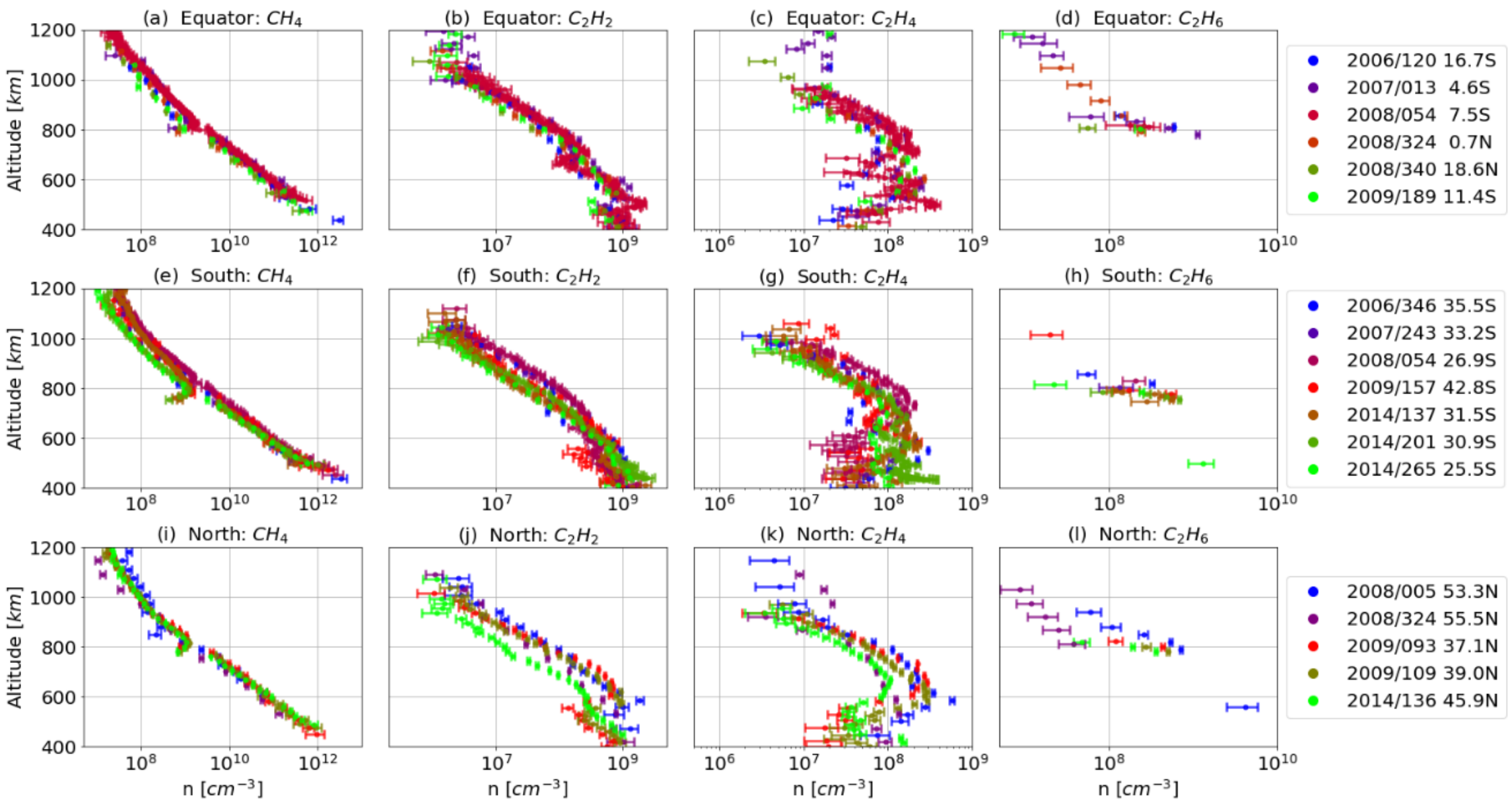

**Figure 9.** Same as Figure 5, but for local densities of species.

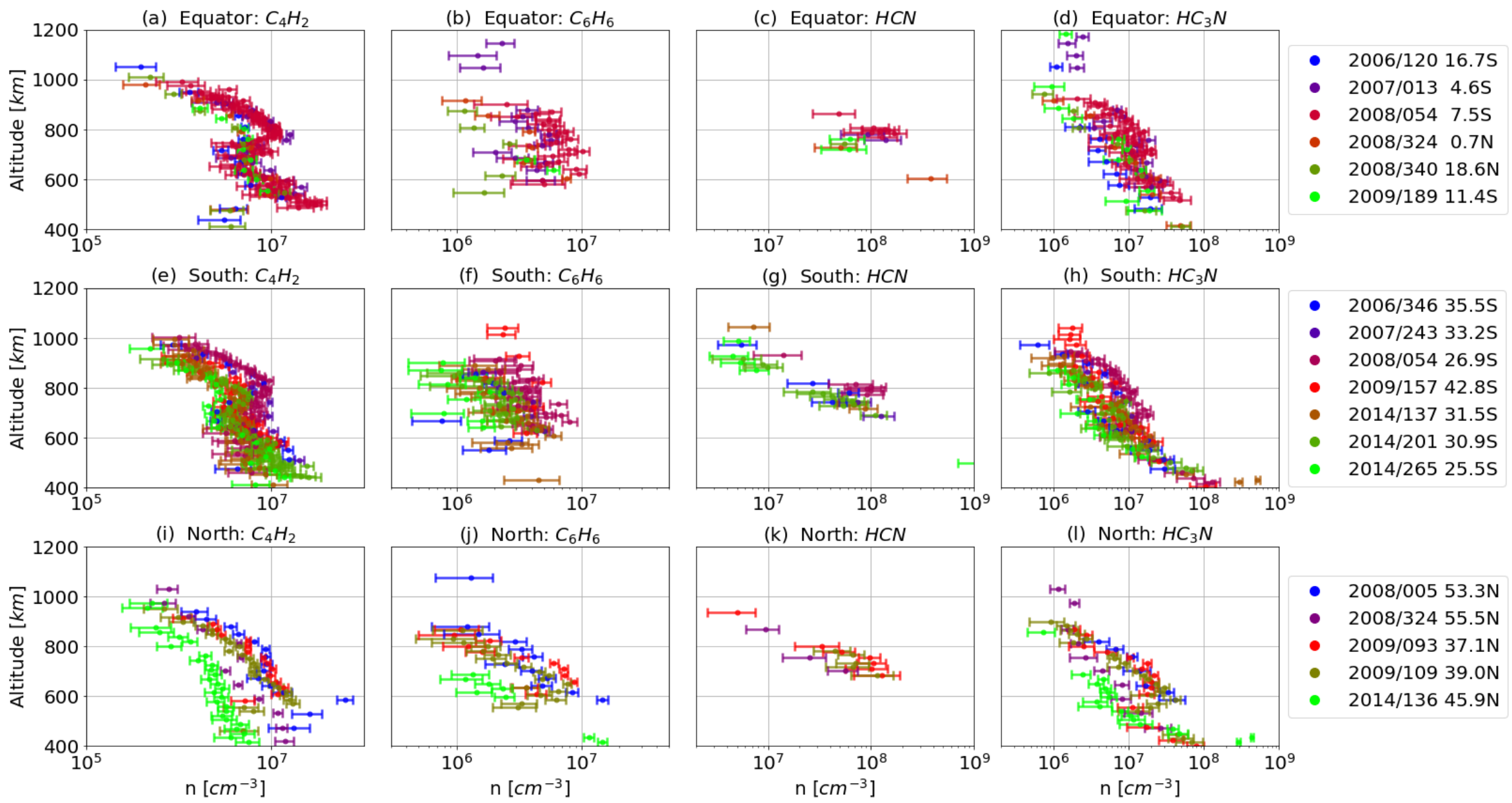

**Figure 10.** Same as Figure 6, but for local densities of species.

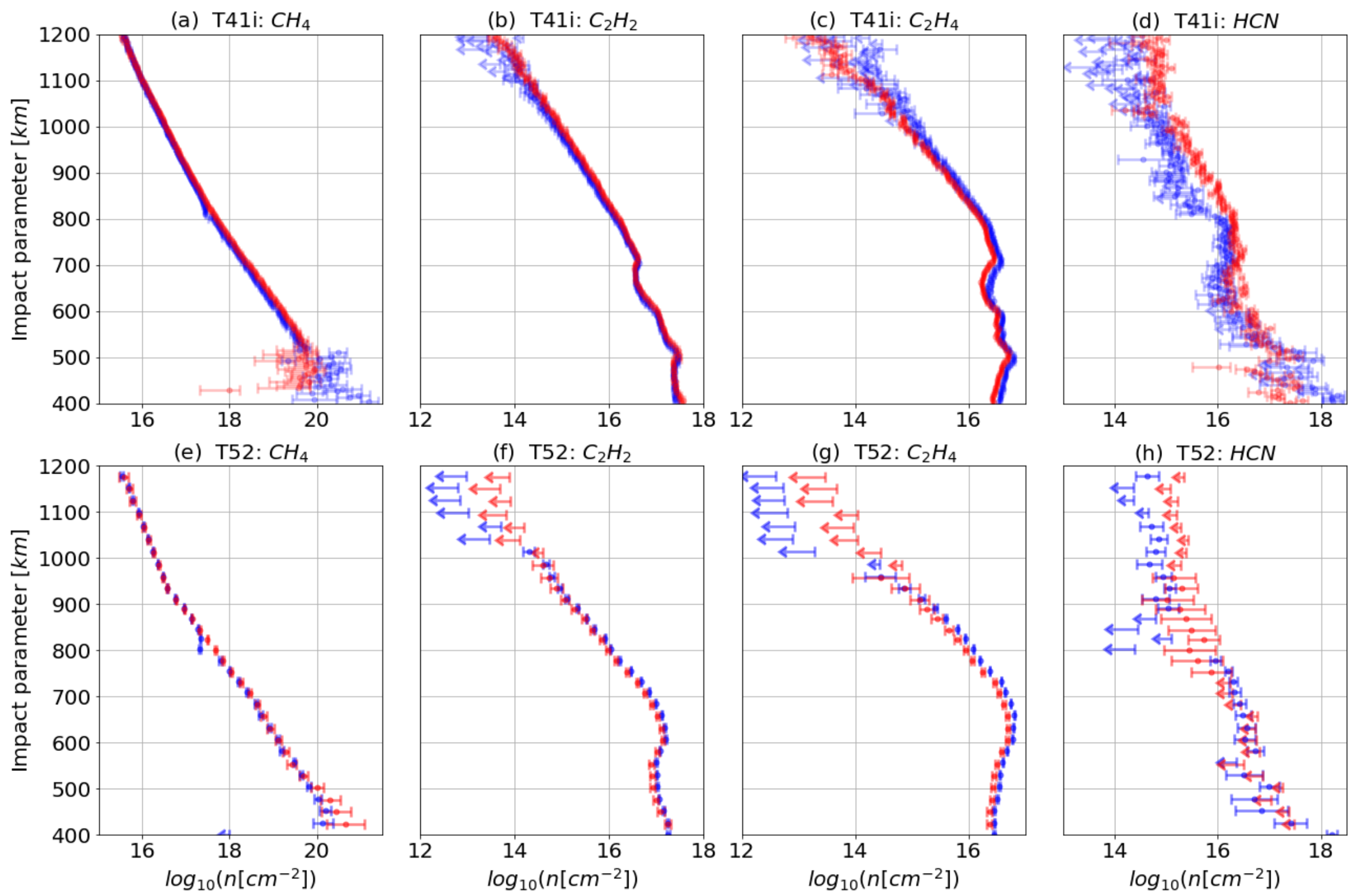

**Figure 11.** (a) LOS abundance of $CH_4$ retrieved from the stellar occultation observation during T41i (blue), compared to Koskinen et al. (2011, red). Dots with error bars denote well-constrained values, while arrows denote upper limits. The lengths of the arrows denote the width of each soft upper limit threshold. (b)-(d), same as (a), but for $C_2H_2$, $C_2H_4$, and HCN, respectively. (e)-(h), same as (a)-(d), but for T52, and for comparison with Fan et al. (2019, red).

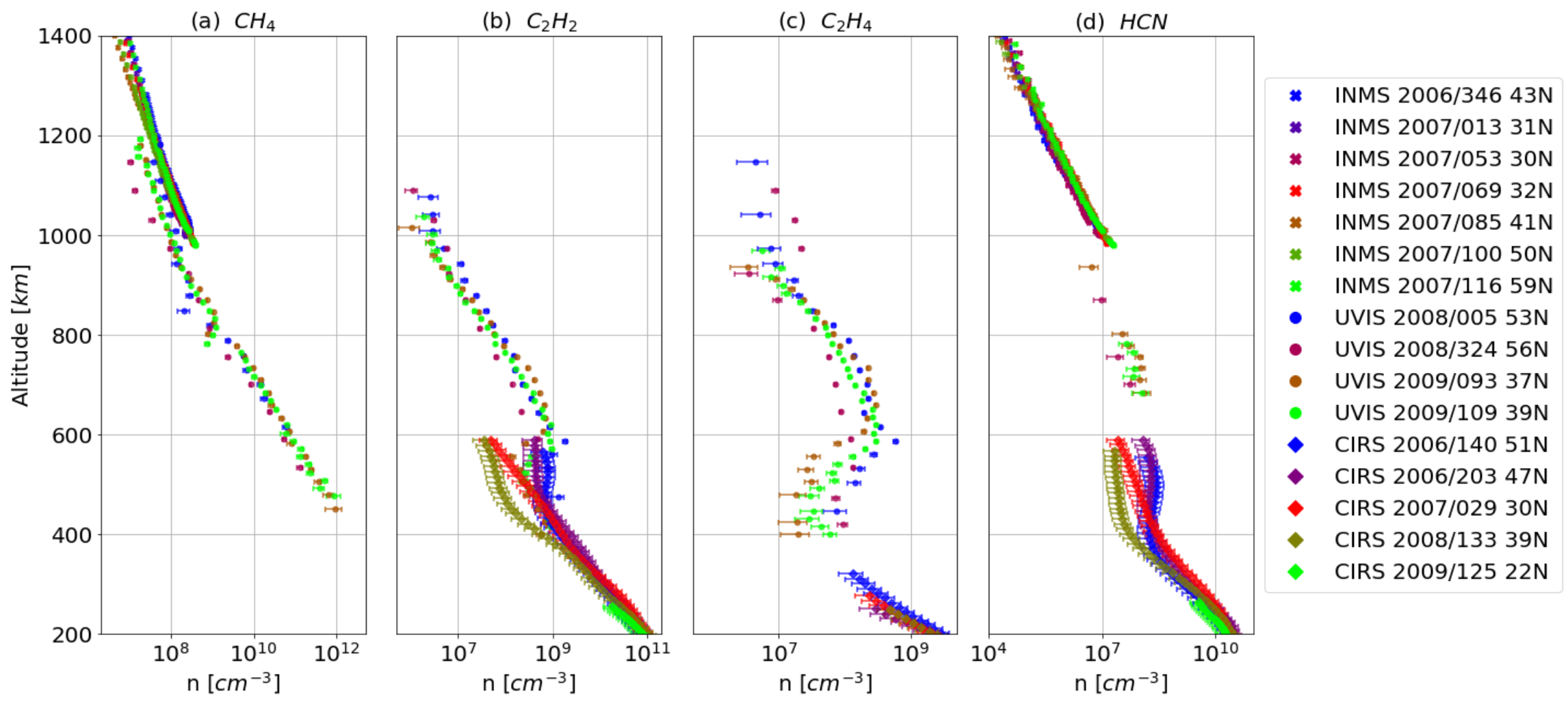

**Figure 12.** (a) Local densities of $CH_4$ retrieved from stellar occultation observations during Titan flybys north of 20°N (Group 3) and before the spring equinox (color dots with error bars), same as those in Figure 9i, compared to results from INMS adapted from Cui et al. (2009, color crosses with error bars) at high altitudes (>1000 km), which includes its instrument recalibration (Teolis et al. 2015). (b) and (c), same as (a), but for $C_2H_2$ and $C_2H_4$, respectively, compared to results from CIRS adapted from Mathé et al. (2020, color diamonds with error bars) at low altitudes (<600 km). (d) Same as (a), but for HCN, compared to both INMS and CIRS results. Time and location of each observation are denoted on the right of the four panels.